\documentstyle[12pt]{article}
\input psfig
\begin{document}
\newcommand{\gsi}{\,\raisebox{-0.13cm}{$\stackrel{\textstyle>}
{\textstyle\sim}$}\,}
\newcommand{\lsi}{\,\raisebox{-0.13cm}{$\stackrel{\textstyle<}
{\textstyle\sim}$}\,}
\newcommand{\be}{\begin{equation}}
\newcommand{\ee}{\end{equation}}
\newcommand{\eq}{\begin{equation}}
\newcommand{\en}{\end{equation}}
\rightline{RU-93-10}
\rightline{CERN-TH.6729/92}
\baselineskip=18pt
\vskip 1.0in

\begin{center}
{\bf \ Baryon Asymmetry of the Universe\\
 in the Minimal Standard Model}
\end{center}

\vskip 0.1in

\begin{center} Glennys R. Farrar\footnote{Research supported in part
by NSF-PHY-91-21039} \\

\vspace{.05in}

{\it Department of Physics and Astronomy \\
Rutgers University, Piscataway, NJ 08855, USA}

\vspace{.25in}

M. E. Shaposhnikov\footnote{On leave of absence from Institute for
Nuclear Research of Russian Academy of Sciences, Moscow 117312,
Russia} \\

\vspace{.05in}
{\it CERN, TH Division \\
CH-1211, Geneva 23, Switzerland}
\end{center}

\vskip 1in

{\bf Abstract:}
We calculate the baryon asymmetry of the Universe which would arise
during a first order electroweak phase transition due to minimal
standard model processes.  It agrees in sign and magnitude with the
observed baryonic excess, for resonable KM parameters and m$_t$ in the
expected range, and plausible values of bubble velocity and other high
temperature effects.

\pagebreak

The existance in the present-day universe of an excess of matter over
antimatter has long been recognized to be a fundamental problem in
cosmology.  A great deal of theoretical work has been done, and a
number of possible explanations have been advanced.  So far, most of
these proposals have required physics beyond the standard model.
Indeed, the existance of the baryon asymmetry of the universe (BAU)
has widely been considered one of the most compelling pieces of
evidence that the standard model is incomplete.

In this Letter we consider the possible production of a baryon number
excess as a result of purely minimal standard model (MSM) processes,
during the electroweak phase transition.  Our calculation is realistic
enough to make it clear that the baryon asymmetry arising from this
mechanism can be responsible for the observed baryon density to
entropy ratio, $n_B/s \sim (4-10)~10^{-11}$, for values of the top
mass and KM parameters in the currently favored ranges.  In order to
settle definitively whether this is the actual origin of the baryonic
asymmetry existing today, it will be necessary to improve our
quantitative understanding of the ew phase transition and of the
behavior of quasiparticles in the high temperature plasma, as well as
refine and extend the analysis reported here.

For an excess of baryons to develop in a Universe which initially has
zero baryon number, the following conditions, first enunciated by
Sakharov, must be met:
\begin{enumerate}
\item  Some interaction of elementary particles must violate
baryon number, since the net baryon number of the universe must
change over time.
\item  C and CP must be violated in order that there is not a perfect
equality between rates of $\Delta {\rm B} \neq 0$ processes, since
otherwise no asymmetry could evolve from an initially symmetric
state.
\item  A departure from thermal equillibrium must play an essential
role, since otherwise CPT would assure compensation between
processes increasing and decreasing the baryon number.
\end{enumerate}
We briefly summarize several features of the standard model which are
necessary to understanding how these requirements will be met:
\begin{enumerate}
\item In the standard model, quarks get their masses as a result of
their Yukawa couplings to the Higgs field.  When the Higgs field has
a non-zero vacuum expectation value (vev), quark masses are induced
which are proportional to their couplings to the Higgs field, times
its vev.  There are off-diagonal Yukawa interactions, in which quarks
of different generations couple to one another through the Higgs
field.  In general the couplings are complex, and for three
generations there is one physically-significant phase, $\delta_{CP}$.
A non-zero value of $\delta_{CP}$ is imagined to be responsible for
the CP violation observed in the kaon system.  The KM matrix
describes the mixing between generations and contains the phase of the
Yukawa couplings.
\item  With three generations the phase in the KM matrix could be
rotated away if any pair of quarks of the same charge were degenerate
in mass, or if any of the mixing angles vanished.  Thus CP violating
effects are significant in particle physics only when a relevant scale
is small enough to be of the order of magnitude of the splitting in
mass between, e.g., $m_s$ and $m_d$.  The $K^o$ system is an example
of this, where the lack of degeneracy between $d$ and $s$ is evidenced
by $m_K \neq m_{\pi}$.  Moreover CP violation in the MSM vanishes
together with
\begin{equation}
{\rm J} \equiv
sin(\theta_{12})sin(\theta_{13}) sin(\theta_{23}) sin(\delta_{CP}),
\label{J_CP}
\end{equation}
using the particle data group convention for the KM mixing angles.
\item Although the standard model Lagrangian conserves baryon number,
quantum effects produce an anomaly which leads\cite{hooft:pr} to
baryon number violating transitions.  While the rate of these
transitions is negligible at $T=0$, at high temperture their rate $\sim
exp [-\frac{\pi M_W(T)}{\alpha_w T}]$\cite{krs85}.  Thus above the
electroweak phase transition the baryon-number-violating transition
rate is rapid compared to the expansion rate of the universe.  We will
require that the vev in the low temperature phase is large enough that
baryon violation is ``turned off", allowing the asymmetry which has
been produced during the transition to survive.  In the MSM where the
only undetermined parameter of the Higgs sector is the Higgs mass,
this requirement may lead (if non-perturbative thermal effects are
unimportant) to an upper limit on the Higgs
mass.\cite{s:m^14}
\end{enumerate}

At temperatures above the electroweak phase transition, the vev of the
Higgs field vanishes\cite{kirlin}; it takes on a constant,
non-vanishing value in the low temperature phase.  We require the
phase transition to be first order, although this depends on the Higgs
mass and is not certain to be true.  During the phase transition,
``bubbles" of the new $< \Phi > \ne 0$ phase nucleate and expand to
fill the universe.  This departure from thermal equillibrium satisfies
the third Sakharov requirement.  Bubble properties, such as their wall
thickness or their expansion velocity, are not well understood at
present.  A number of phenomena could produce a baryon excess during
the ew phase transition.  For definiteness we concentrate on a
specific mechanism; see \cite{fs:2} for others.

As a result of their thermal motion, quarks and antiquarks in the
neighborhood of the bubble wall propagate through it.  Since their
masses result from their interaction with the vev, they see the bubble
wall as a potential barrier and scatter from it.  We model this
process in detail\cite{fs:2}, keeping the plasma masses of the quarks and
antiquarks which originate from their interactions with the gauge and
Higgs particles present in the heat bath, as well as treating quantum
mechanically the process of their interaction with the bubble wall of
Higgs field.  Due to the spatial variation of the effective CP
violating phase, which comes about because the physical eigenstates
depend on the interplay between flavor dependent thermal effects and
the masses induced by the changing Higgs vev, there can be a
difference between the reflection and transmission coefficients of
quarks and antiquarks.  We report below on our computation of this
asymmetry, in the one-dimensional problem which results when quark
motion parallel to the bubble wall is ignored.  Due to the fact that
Lorentz invariance is broken by the thermal medium, momentum
components parallel to the wall could be dynamically important, but
that issue will be left for future work.

The total baryonic current is conserved in quark scattering with the
bubble wall.  Nonetheless, if there is an asymmetry in reflection and
transmission coefficients the wall would separate particles and
antiparticles, with, e.g., quarks flowing preferentially toward the
low-temperature phase and antiparticles toward the high-temperature
side.  In the high temperature phase, sphaleron transitions operate to
equilibrate the antiquark excess\cite{krs85}, converting most of them
to quarks and leptons.  But as long as the vev in the low temperature
phase is sufficiently large, the sphaleron transition rate is too low
to keep up with the expansion rate of the universe, and the quark
excess is preserved until now.\footnote{ The idea that the BAU could
result from the separation of a quantum number by the bubble wall
combined with equilibration of this quantum number in the high
temperature phase due to baryon-number-violating sphaleron processes,
originated with Cohen, Kaplan and Nelson, employing a lepton-number
violating interaction with the bubble wall\cite{ckn:L1}. The idea that
MSM interactions of quarks and antiquarks with the bubble wall could
directly cause a separation of baryon number is due to MS\cite{s:msm},
where the element of including thermal effects is also introduced.}

As noted above, if two same-sign-quarks are degenerate in mass there
is no CP violation, since the phase $\delta_{CP}$ can be removed from
the KM matrix by a change in the definition of the overall phases of
quark fields.  This fact manifests itself in the present context by a
tendency for different flavor eigenstates to have canceling
contributions to the baryonic asymmetry current.  For instance, if the
reflection probability at a given energy for a $d_L$ is greater than
that for an anti-${d_L}$, the reflection probability of an $s_L$ will
be less than that for an $\bar{s_L}$ by a nearly identical amount.  In
the limit $m_s - m_d \rightarrow 0$ the compensation is perfect, when
reflection involving $b$'s is included.  An estimate of the residual
CP violating asymmetry for typical quark energy $\sim T$ is
$ (m_t^2 - m_c^2)(m_t^2 - m_u^2)(m_c^2 - m_u^2)(m_b^2 - m_s^2)(m_b^2 -
m_d^2)(m_s^2 - m_d^2)/T^{12}$ times J, $\sim 10^{-21}$\cite{s:m^14}.
For this reason it has commonly been believed that the MSM cannot
be responsible for the BAU.

The new observation of the present work is that the important quark
energies are $not~\sim T$, but rather energies such that the $s$ quark
is totally reflected but the $d$ quark is partly reflected and partly
transmitted, maximizing the dynamical difference in their
contributions to the asymmetry current.  Taking into account thermal
effects on the quark propagation (see below and ref. \cite{fs:2}), one
finds that the $p_t$ for which $s$ quarks are reflected is $p_t \sim
\frac{9\alpha_W}{32}\sqrt{\frac{3\pi}{2\alpha_s}} T \sim 10^{-1}T$.  The
range of $p_t$ with $s$ but not $d$ reflection is $\sim {{m_s -
m_d}\over T}$.

The net baryonic flux through the wall is proportional
to the group velocity of the quasi-particles perpendicular to the wall
($\sim \frac{1}{3}$) and to the asymmetry in Fermi-distributions on
the two sides of the wall due to its motion with respect to the themal
medium ($\sim 2v\frac{\partial \omega}{\partial
p_t}\frac{\omega}{T}\sim\frac{2}{3}v \sqrt{\frac{2 \pi\alpha_s}{3}}$
for small wall velocity,$v$, where $\omega = \omega(p)$ is the
quasiparticle dispersion relation).  The strength of CP violation in
this region is just J, given in eqn (\ref{J_CP}).  Thus putting all
the factors together we expect
\begin{equation}
n_b/s \sim \frac{2}{9} \sqrt{\frac{2 \pi \alpha_s}{3}}\left(\frac{m_s
- m_d}{T}\right)\frac{{\rm J}~v}{N_{eff}},
\end{equation}
where $N_{eff} \sim 100$ is the total number of degrees of freedom.
Global fits to determine KM parameters place J in the
range\cite{gn} $(1.4-5.0)10^{-5}$, so
we expect $n_b/s\sim (2-8)10^{-11}v$, which can be of the right
order-of-magnitude to account for the BAU.

Having outlined the way in which CP violation in the quark mixing
matrix can lead to a present-day baryon asymmetry, we next calculate
it more quantitatively, in the quark-antiquark separation mechanism.
There are two essential effects to be included: The interactions of
the quarks and antiquarks with the plasma of gluons, ew gauge bosons
and Higges, and the quantum mechanical scattering of the quarks and
antiquarks from the bubble wall in the Higgs vev.  The effects of the
interactions of the quarks and antiquarks with the gauge and Higgs
particles in the heat bath are most efficiently taken into account by
changing variables to a quasi-particle description.  The propagators
of the quasi-particles have been determined to one-loop accuracy,
neglecting internal masses, by Klimov\cite{klimov} and
Weldon\cite{weldon}.  The quasiparticles have interesting and
unfamiliar behavior, but space limitations prevent us from describing
them further here.  These and other details are given in a longer
article\cite{fs:2}.  Here we record only the equation of motion which
these quasiparticles obey in the wall rest frame, in the limit which
is relevant to total reflection of the strange quark, momentum $\ll
\omega$, treating only the motion normal to the bubble wall, and in
the limit of small plasma velocity with respect to the wall:
\begin{equation}
\pmatrix{\omega(1 + \alpha_L + \beta_L) + i \frac{\partial}{\partial
x} (1 +
\alpha_L) & KM_d(x) \cr
(KM_d(x))^{\dagger} & \omega(1 + \alpha_R + \beta_R) -
i\frac{\partial}{\partial x} (1 + \alpha_R) \cr}
\pmatrix{L \cr
R \cr} = 0 \label{eq:eom}
\end{equation}
where $\alpha_{L,R}, \; \beta_{L,R}$ and $M_d(x)$ are 3 $\times$ 3
matrices and $L$ and $R$ are 3 component spinors in flavor-space.
$M_d(x)$ is the Higgs-induced mass at $T_c$.  For charge -1/3 quarks
in the unbroken phase, in the gauge basis where the interactions of
quarks with the W and Z are diagonal:
\begin{eqnarray}
\alpha_{L} &=& - [\frac{4 \pi}{3} \; \frac{\alpha_s}{6} + \frac{4
\pi}{3} \; \frac{\alpha_w}{32} (3 + \frac{m^2_{u}}{M_w^2} +
\frac{{K m^2_{d}K^{\dagger}}}{M_w^2} +
\frac{\sin^2 \theta_w}{9})]\frac{T^2}{\omega^2} \nonumber \\
\alpha_{R} &=& - [\frac{4 \pi}{3} \; \frac{\alpha_s}{6} + \frac{4
\pi}{3} \; \frac{\alpha_w}{32} (\frac{m_u^2}{M_w^2} +
\frac{16}{9} \sin^2 \theta_w)]\frac{T^2}{\omega^2},
\label{eq:alphaLR}
\end{eqnarray}
where $m^2_u$ and $m^2_d$ are diagonal matrices of the charge +2/3 and
-1/3 masses at $T = 0$, and $K$ is the KM matrix.  For momenta small
compared to $\omega$, $\beta_{L,R} = 2 \alpha_{L,R}$.  More general
expressions can be found in \cite{fs:2}.

We can solve these equations analytically in two limits: no mixing and
zero wall thickness with small mixing.  For more realistic cases they
must be solved numerically, although having the exact cases to verify
the correctness and accuracy of our numerical solutions is very
useful.  Details of the analytical and numerical results are given in
ref. \cite{fs:2}.  We find that when the energy is such that neither
or both $d$ and $s$ quarks are totally reflected, the difference
between reflection probabilities of quarks and antiquarks, after
summing over all three flavors, is extremely small.  (Less than our
numerical integration accuracy of one part in $10^{10}$.)  However, as
expected on the basis of the heuristic discussion above, the asymmetry
is substantial in the narrow energy window in which the $s$ quark is
totally reflected but the $d$ is not.

The figure shows $\Delta$, the difference in the reflection
probabilities for right chiral quarks and antiquarks incident from the
unbroken phase, summed over flavors, in the interesting range of
energies.  We take as representative values $m_t = 150 \rm{GeV}$, $m_c
= 1.6\rm{GeV}$, $m_u = .005 \rm{GeV}$, $m_b = 5 \rm{GeV}$, $m_s =
.15\rm{GeV}$, $m_d = .01 \rm{GeV}$, $ s_{12} = .22, s_{23} = .05,
s_{13} = .007$ and $sin(\delta_{CP}) = 1$.  For the calculation of
this figure the wall velocity, $v$, was zero and the wall thickness
was 10/T$_c$, a popular value.  Taking the wall to be narrower does
not significantly change the result; taking it a factor of three
thicker increases the result by a factor of two.  This dependence on
wall thickness is not surprising: even in the thin wall limit there is
a non-trivial CP conserving phase shift to interfere with the KM CP
violating phase.  The asymmetry in the reflection probabilities
increases when the effect of the flow of the thermal medium is
included\cite{fs:2}, so that $\Delta$ is almost a factor of five
larger for $v=.25$ than it is for the $v=0$ case shown in the figure.

The upper pair of peaks occupy the energy range in which the strange
quark is totally reflected.  Note that the width in energy of this
region is $\sim .1$ GeV, just the mass of the strange quark at
that temperature. The ``notch" in the middle, of width $\sim .006 \sim
m_d(T)$, is the region in which the down quark is also totally
reflected and GIM cancellation is perfect, as expected.  The
unfamiliar feature that total reflection occurs for a range of
energies, rather than for all energies less than some value, results
from the unusual properties of the quasi-particle dispersion
relation\cite{fs:2}, but is not essential to our result. Our analytic
calculation in thin-wall, small mixing approximation\cite{fs:2}
provides an adequate description of reflection in this region.  At
lower energy there is another region of a different character,
involving level crossing between $d$ or $s$ and $b_R$.  It would not
be present if mixing were absent. Its width is $\sim m_b(T)\sin(\theta_{23})$.

We have checked that the asymmetry vanishes as pairs of masses are
brought together.  When $m_s \rightarrow m_d$ it arises by the
squeezing away of the width of the upper peaks and the diminution of
the magnitude of the lower peak.  When masses in the charge +2/3
sector are brought together, or $m_b \rightarrow m_{d,s}$, the
magnitudes of the peaks decrease appropriately.  We checked that the
result vanishes as mixing angles are taken to zero, although in the
physical range of $\theta_{23}$ the result is non-linear, increasing
by 40\% as $\theta_{23}$ is changed from 0.05 to 0.06, and changing
sign for $\theta_{23} \lsi 0.03$.  The $m_t$ dependence is interesting:
for low values, $m_t \lsi 110$GeV, the integrated asymmetry has the
opposite sign as for $m_t \gsi 110$GeV.  It reaches its maximum value
for $m_t \sim 210$GeV, where it is more than four times greater than
for $m_t=150$GeV, then decreases for larger $m_t$.  For a more
detailed discussion and additional figures see ref. \cite{fs:2}.

The net $L$ baryonic current resulting from the asymmetry in
reflection coefficients for $R$'s incident from the unbroken phase and
$L$'s incident from the broken phase is obtained from $\Delta$ as
follows\cite{fs:2}:
\[
J^L_{CP} = \int \frac{ d \omega}{2 \pi} \left( n_F[\omega(1 -
\frac{d\omega}{dp} v)] - n_F[\omega(1 + \frac{d\omega}{dp} v)]
\right)\Delta(\omega).
\]
Given the net baryonic current flowing through the bubble wall,
corresponding to a preferential flow of baryons in one direction and
antibaryons in the other, we next wish to determine the resultant
$n_B/s$, assuming sphaleron transitions operate on $L$ chiral quarks
and antiquarks to balance the chemical potentials in the high
temperature phase, but are completely suppressed in the low
temperature phase.  To make an accurate estimate requires a detailed
study of the non-equillibrium statistical physics of the problem,
however we can obtain a conservative estimate of $n_B/s$ as follows.
Suppose that the wall velocity is low, so that diffusion permits a
back-current of baryon number to be established, which acts to replace
the anti-baryon number which is being destroyed in the high
temperature phase at a rate $\Gamma$ by the sphaleron transitions.
For sufficiently low velocity the problem is essentially static and the
result is \cite{fs:2} $n_B = 3 J_{CP}^L f(\rho)$,
where $\rho = {3D \Gamma \over v^2}$. $f(\rho) = 1$ for $\rho \gg 1$
and $f(\rho) = \frac{4}{9}\rho$ for $\rho \ll 1$.  The physical
importance of $\rho$ is clear since the typical distance from the
wall, of a particle scattered at $t=0$ into the unbroken phase, is
$\sqrt{Dt}-vt$. Thus $\frac{D}{v^2}$ is the typical time in which
the sphaleron transitions can act on that particle before it is
enveloped by the expanding low-temperature phase.  The correct values
to take for the diffusion length, $D$, and sphaleron rate and wall
velocities are very uncertain, but can plausibly be such that
$f(\rho)\sim 1$, though a suppression as large as $10^{-3}$ is also
possible \cite{fs:2}.

Since we have computed the current in one-dimension, we divide by the
one-dimensional entropy for the MSM particle content at the ew phase
transition: $s_{1-d}=\frac{73 \pi T}{3}$.  Taking $v=.25$ values for
reflection probabilities and the Boltzman factors, we find
\begin{equation}
n_B/s \sim 4~ 10^{-11} \left(\frac{{\rm J}}{.22 \times .05 \times
.007}\right)~f(\rho),
\label{asymfinal}
\end{equation}
for $m_t=150$GeV, $\theta_{23}=0.05$, and inverse wall thickness
10GeV.  (Until the result of the full three-dimensional calculation is known,
one should take this result as an order-of-magnitude estimate.)  Since
the prediction increases rapidly for larger values of $m_t$ and wall
velocity, there seems to be ample margin within the favored ranges of
these quantities\footnote{Recent estimates place $.1 < v < 1$, see
\cite{fs:2} for refs.} to tolerate some suppression from the uncertain
overall factor $\sin \delta_{CP} (\frac{\sin
\theta_{13}}{.007})f(\rho)$ and the difference between the one- and
three-dimensional cases.\footnote{Of course if there were a fourth
generation with a comparable KM structure then the degeneracy between
$d$ and $s$ would be irrelevant and would be replaced by the
degeneracy between $b$ and $s$.  Using the analytic thin wall
expression\cite{fs:2} yields an enhancement factor
$\sim(\frac{m_{t'}}{m_t})^4(\frac{m_{t}}{m_c})^2(\frac{m_{b}}
{m_{b'}})^2\frac{s_{23}s_{24}s_{34}}{s_{12}s_{23}s_{13}},~\sim10^3$ for
$m_{b'}\sim m_t$ and $m_{t'}\sim 500$GeV.}

While the sign of $\sin \delta_{CP}$ is not at present unambigously
determined\cite{nq}, a positive sign is favored.  In this case
(\ref{asymfinal}) correctly predicts a baryon, not antibaryon, excess.
Within the framework of this model, effects which we have not included
seem capable only of affecting the magnitude and not the sign of this
prediction (see ref. \cite{fs:2}).  Changes in $v$ and wall thickness,
and changes of the poorly known $m_t$, $\theta_{23}$, and
$\theta_{13}$ within their favored ranges, do not change the sign of
(\ref{asymfinal}).  Thus refinements in the treatment of this problem
may not modify the conclusion, that minimal standard model
interactions can be responsible for the BAU.

The most crucial outstanding problems are those associated with our
still-primitive ability to deal with the high temperature environment
during the electroweak phase transition: sphaleron rate, wall
velocity, quasiparticle scattering length are obvious examples.  In
addition, our calculation of the quasi-particle
reflection coefficients requires knowing their equation of motion in
the plasma, which we determined by using propagators calculated to
1-loop accuracy in high temperature perturbation theory.  Until plasma
screening and ``confinement'' effects are better understood, these
propagators must considered as approximate.  Also, a more complete calculation
would include incoherent as well as coherent processes occuring during
the quantum mechanical scattering from the vev.

To summarize, we have argued that already-known physics of the minimal
standard model may explain the observed baryon asymmetry of the
universe.  A quantitative calculation in a specific mechanism gives
the correct sign and magnitude.  The essential new ingredient is not
overlooking those regimes of quark momenta in
which the most degenerate pair of quarks have very different dynamical
behavior.  If this is the explanation for the baryon asymmetry of the
universe, then future precision comparisons between observation
and theory will provide a powerful test of our understanding of the ew
phase transition, as well as constrain the KM matrix and the masses and
generation content of the standard model.

{\bf Acknowledgement:}
GRF is grateful for the support and hospitality of CERN during a
sabbatical visit in which most of this work was done, and MES thanks
Rutgers University for its kind hospitality during the completion of
the work.  We have benefitted from many helpful discussions with
colleagues at both institutions.  A. Terrano made particularly
significant contributions.

\psfig {file=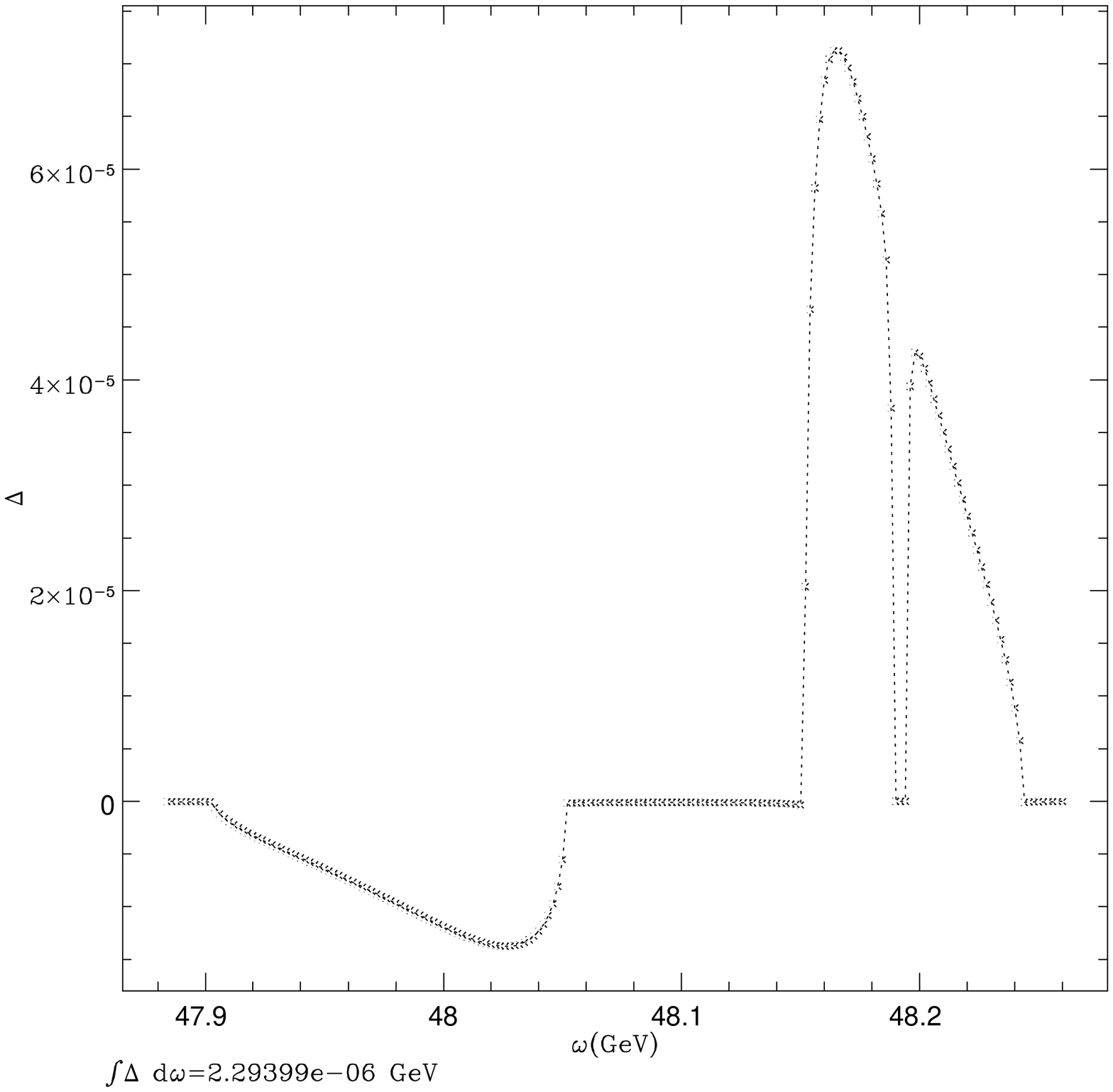,width=6.5in}

{\bf Figure Caption:} $\Delta$, the asymmetry in the reflection
probabilities of right-chiral quarks and antiquarks incident from the
unbroken phase, for zero wall velocity.


\begin{thebibliography}{10}

\bibitem{hooft:pr}
G.~'t~Hooft.
\newblock {\em Phys. Rev.}, D14:3432, 1976.

\bibitem{krs85}
V.~A. Kuzmin, V.~A. Rubakov, and M.~E. Shaposhnikov.
\newblock {\em Phys. Lett.}, 155B:36, 1985.

\bibitem{s:m^14}
M.~E. Shaposhnikov.
\newblock {\em JETP Lett.}, 44:465, 1986.

\bibitem{kirlin}
D.A.Kirzhnitz and A.D.Linde.
\newblock {\em Phys. Lett.}, 72B:471, 1972.

\bibitem{fs:2}
G.~R. Farrar and M.~E. Shaposhnikov.
\newblock Technical Report RU-93-11, CERN-TH.6732/92, Rutgers/CERN, 1993.

\bibitem{ckn:L1}
A.~Cohen, D.~Kaplan, and A.~Nelson.
\newblock {\em Phys. Lett.}, B245:561, 1990.

\bibitem{s:msm}
M.~E. Shaposhnikov.
\newblock {\em Phys. Lett.}, B277:324, 1992;B282:483(E),1992.

\bibitem{klimov}
V.~V. Klimov.
\newblock {\em Sov. J. Nucl. Phys.}, 33:934, 1981.

\bibitem{gn}
Y.~Grossmann and Y.~Nir.
\newblock Technical Report in preparation., Weizmann Institute, 1993.

\bibitem{weldon}
H.~A. Weldon.
\newblock {\em Phys. Rev.}, D26:2789, 1982.

\bibitem{nq}
Y.~Nir and H.~Quinn.
\newblock {\em Phys. Rev.}, D42:1473, 1990.

\end{thebibliography}
\end{document}